\documentclass[10pt,twocolumn,aps,american,prl,showpacs,preprintnumbers,amsmath,amssymb]{revtex4-2}

\usepackage{babel} 	
\usepackage[utf8]{inputenc}
\usepackage{lineno,hyperref}
\usepackage{amsbsy}
\usepackage{amstext}
\usepackage{amssymb}
\usepackage{graphicx}
\usepackage{esint}
\usepackage{amsmath}
\usepackage[all]{xy}
\usepackage{multirow,array}
\usepackage{xcolor}
\usepackage{float}
\usepackage{units}
\usepackage{multirow}
\usepackage{array}
\usepackage[titletoc]{appendix}
\usepackage{rotating}
\usepackage{dcolumn}
\usepackage{bm}
\usepackage{fancyhdr} 
\usepackage{braket}						
\usepackage{color} 							
\usepackage{lastpage} 						
\usepackage{hyperref} 						
\usepackage{url}
\usepackage{bibentry}
\usepackage{multirow}
\usepackage{silence}
\usepackage{appendix}
\usepackage{natbib}
\usepackage{physics}

\usepackage{tikz}
\usetikzlibrary{shapes}

\begin{document}
\renewcommand{\figurename}{FIG.}
\title{Quantum Lyapunov exponent in dissipative systems}

\author{Pablo D. Bergamasco}
\affiliation{Departamento de F\'isica, CNEA, Libertador 8250, (C1429BNP) Buenos Aires, Argentina}
\author{Gabriel G. Carlo}
\author{Alejandro M. F. Rivas}
\affiliation{Departamento de F\'isica, CNEA, CONICET, Libertador 8250, (C1429BNP) Buenos Aires, Argentina}

\date{\today}

\begin{abstract}
The out-of-time order correlator (OTOC) has been widely studied in closed quantum systems. However, there are very few studies for open systems and they are mainly focused on isolating the effects of scrambling from those of decoherence. Adopting a different point of view, we study the interplay between these two processes. This proves crucial in order to explain the OTOC behavior when a phase space contracting dissipation is present, ubiquitous not only in real life quantum devices but in the dynamical systems area. The OTOC decay rate is closely related to the classical Lyapunov exponent --with some differences -- and more sensitive in order to distinguish the chaotic from the regular behavior than other measures. On the other hand, it reveals as a generally simple function of the longest lived eigenvalues of the quantum evolution operator. We find no simple connection with the Ruelle-Pollicott resonances, but by adding Gaussian noise of $\hbar_{\text{eff}}$ size to the classical system we recover the OTOC decay rate, being this a consequence of the correspondence principle put forward in  [\href{https://journals.aps.org/prl/abstract/10.1103/PhysRevLett.108.210605}{Physical Review Letters 108 210605 (2012)} and \href{https://journals.aps.org/pre/abstract/10.1103/PhysRevE.99.042214}{Physical Review E 99 042214 (2019)}]
\end{abstract}

\maketitle
The out-of-time order correlator (OTOC) has been initially introduced in the context of superconductivity, where exponential growth as a function of time has been associated with the behavior of chaotic \cite{larkin1969quasiclassical} systems.  Pioneering work in black hole theory \cite{maldacena2016bound} has led to a resurgence of interest in this measure in various fields such as many-body physics \cite{shenker2014black, aleiner2016microscopic,huang2017out,borgonovi2018emergence,slagle2017out,chen2017out,garcia20}, high energy theory \cite{akutagawa2020out} and quantum chaos \cite{lakshminarayan2018out, wang2021quantum,jalabert2018semiclassical}

OTOC is conceptually related to quantum information scrambling and complexity \cite{pappalardi2018scrambling,campisi2017thermodynamics,swingle2018unscrambling, balachandran2021eigenstate, PhysRevResearch.2.043178, prakash2019scrambling}. For weakly coupled strongly chaotic systems, the OTOC has been found to initially grow in a regime related to intra-subsystem scrambling, while during a second regime it approaches saturation depending on the coupling interaction \cite{prakash2019scrambling}.
On the other hand, OTOC is a good indicator of quantum complexity \cite{PhysRevResearch.1.033044, bergamasco2017,Benenti-Carlo-Prosen}. In this line, an equivalence between the average OTOC over a full basis of operators and the linear entropy has been established through the OTOC-Renyi theorem \cite{hosur2016chaos, fan2017out}; hence, its evaluation allows to distinguish between regular and chaotic behavior.  This equivalence can be obtained in a restricted set of meaningful operators \cite{bergamasco2020relevant} which amounts to defining a preferred OTOC basis.

All these results have been obtained for closed systems, but notably during the last couple of years it appeared a strong interest in the OTOC behavior for open systems. 
Dissipation and information scrambling interplay has been found to manifest in the details of the decay of the 4-point out-of-time ordered correlator of local operators \cite{zhang2019information}. Due to the information loss induced by the environment, the OTOC has been considered not capable of distinguishing between scrambling and decoherence \cite{touil2020quantum}. Instead, mutual information has been proposed as a better quantifier of these contributions to the information flow \cite{touil2021information}. It has only recently been shown that the interplay between information scrambling and decoherence can be used to differentiate between chaotic and regular regimes in dissipative many-body spin chains \cite{zanardi2021information}, but only for bipartite OTOC and in an inherently quantum setting. 

In the context of open quantum systems with a clear classical counterpart, is the OTOC capable of distinguishing between regular or chaotic behavior, as the classical Lyapunov does?
In this letter we study the OTOC for generic systems of this kind, understood as the quantized versions of typical classical dynamical systems (overwhelmingly studied in the nonlinear dynamics area), by considering the paradigmatic dissipative modified kicked rotator map (DMKRM). For this system and the parameter values explored, we observe that there is no short times Lyapunov exponential growth. However, at longer times ($t > 5$) the OTOC decays exponentially at a rate that closely follows the Lyapunov exponent, regardless of the operators or the initial state. This decay depends on the forcing and dissipation strengths ($K$ and $\gamma$ parameters respectively) which 
set the degree of scrambling and environmental effects. The interplay between these two processes determines the OTOC decay regime, quantitatively related to the details of the spectral gap of the quantum evolution operator. 
It is worth mentioning here that the statistical properties of dynamical systems at the classical level change across local bifurcations, where neither the Lyapunov exponents nor the covariant Lyapunov vectors provide a good criterion for determining the stability of the attractors. Instead, the decay of correlations or mixing, which can be estimated from long times series, are related to  
the eigenvalues of the generator of the transfer operator semigroup \cite{tantet2018resonances}.

The OTOC is usually defined as:
\begin{align}
    C(t) = \expval{[\hat{A}(0),\hat{B}(t)][\hat{A}(0),\hat{B}(t)]^{\dagger}}
    \label{eq:otoc}
\end{align}
where $\hat{A}$ and $\hat{B}$ are two operators at different times evolved in the Heisenberg picture. The mean value is taken over a localized initial state $\expval{\cdots}=\bra{\psi_{0}} \dots \ket{\psi_{0}}$. Nevertheless, the results are the same in the case of considering a thermal initial state \cite{garcia2018chaos}. Our model consists of a particle moving in one dimension subjected to a periodically kicked asymmetric potential,     
\begin{equation}
    V(q,t) = k \left[\cos{(q)} + \frac{a}{2}\cos{(2q+\phi)}\right]\sum_{m=-\infty}^{\infty}{\delta(t - m\tau)}
    \label{eq:potencial}
\end{equation}
Where $k$ is the strength of the kick and $\tau$ its period. Adding dissipation, we get \cite{carlo2005}
\vspace{-2mm}
\begin{equation}
    \Bar{n} = \gamma n + k[\sin{(q)} + \sin{(2q+\phi)}]\qquad  \Bar{q} = q + \tau \Bar{n}
    \label{eq:krmap}
\end{equation}
where $n$ is the momentum variable conjugate to $q$ and $\gamma$ $(0 \leq \gamma \leq 1)$ is the dissipation parameter. With $\gamma = 1$ we recover the conservative system, while setting $\gamma=0$ corresponds to maximum environmental strength. Usually, a scaled momentum $p=\tau n$ and the quantity $K=\tau k$ are introduced in order to simplify the expressions. We take $a=0.5$ and $\phi=\pi/2$ (these parameters are related to spatio-temporal symmetries and at these values provide with a rich dynamical landscape suitable for our investigation). 

In order to quantize the model, we follow the standard procedure $q \rightarrow \hat{q} $ and $n \rightarrow \hat{n} = -i(d/dq)$ $(\hbar = 1)$. Given that $[\hat{q},\hat{p}]=i\tau$ (where $\hat{p}=\hat{n})$, we define the effective Planck constant by means of identifying $\hbar_{\text{eff}}=\tau$. In the classical limit, $\hbar_{\text{eff}} \rightarrow 0$ and $K=\hbar_{\text{eff}} k$ remains constant. On the other hand, dissipation is treated in the usual way through the Lindblad master equation \cite{lindblad1976} to describe the evolution of operators in the Heisenberg representation,
\begin{align}
\Dot{\hat{B}} = i[\hat{H}_{s}, \hat{B}] - \frac{1}{2}\sum_{\nu=1}^{2}\{\hat{L}^{\dagger}_{\nu}\hat{L}_{\nu}, \hat{B}\} +\sum_{\nu=1}^{2} \hat{L}^{\dagger}_{\nu} \hat{B} \hat{L}_{\nu} \equiv \mathcal{L}(\hat{B})
\label{eq:lindblad_B}
\end{align}
where ${H}_{s}={H}^{2}/2 + V(\hat{q},t)$ is the Hamiltonian of the system, $\{\ ,\ \}$ the anticommutator, and $\hat{L}_{\nu}$ are the Lindblad operators defined as \cite{carlo2019}:
\begin{equation}
    \begin{gathered}
        \hat{L}_{1} = g \sum_{n} \sqrt{n+1}\ket{n}\bra{n+1}\\
        \hat{L}_{2} = g \sum_{n} \sqrt{n+1}\ket{-n}\bra{-n-1}
    \end{gathered}
\end{equation}
where $ \ket{n} $ are the momentum states with $ n=0, 1, \dots $ and $ g = \sqrt{-\ln\gamma} $ \cite{graham1985, dittrich1989}.

In the following we compare the OTOC behavior in several areas of the parameter space (corresponding to different dynamical regimes), both with the inverse participation ratio (IPR) (another widely used quantum complexity measure) and 
the (classical) Lyapunov exponent which allows distinguishing between regular and chaotic behavior efficiently. We use the operators $\hat{A} = e^{i\hat{Q}}$ and $ \hat { B } = \hat{ P } $ in Eq.\eqref{eq:otoc}, where $ \hat{ Q } $ is the position operator and $ \hat{ P } $ is the momentum operator. The evolution of 
$ \hat{ B } $ is given by Eq.\eqref{eq:lindblad_B} integrated between $ t $ and $ t+1 $, i.e. we are looking at the state of the system between successive potential kicks. As initial condition, we have used a coherent state centered at $ \expval{ p_{0} } = 0 $ and $ \expval{ q_{0} } = \pi $. In the classical case, we numerically evaluate Eq. \eqref{eq:krmap}. Our results have shown to be robust against a change in the operators $ \hat { A } $, $ \hat { B } $ considered for the calculations. 

\begin{figure}
    \centering
    \includegraphics[width=7.0cm, angle=-90]{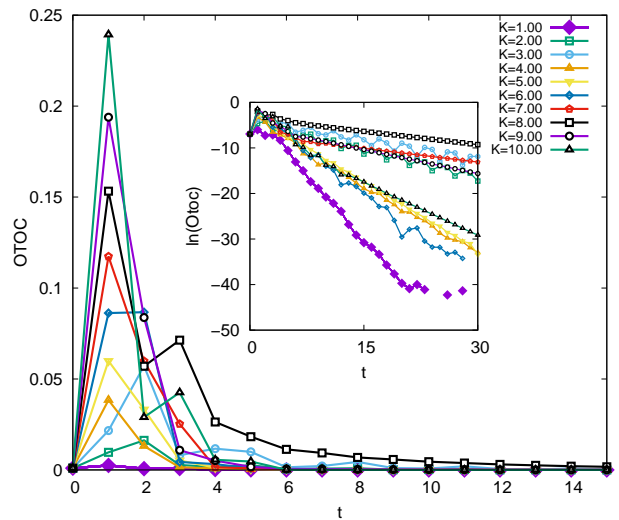}
    \caption{(Color online) OTOC as a function of time $t$ for different values of $ K $ represented by lines with symbols (see legend). 
    The inset shows the same evolution in log scale. In this case $ \hbar_{\text{eff}} = 0.031 $, $ \gamma = 0.200 $ and $ N = 1024 $.}
    \label{fig:fig1}
\end{figure}
\begin{figure}
    \centering
    \includegraphics[width=7.0cm, angle=-90]{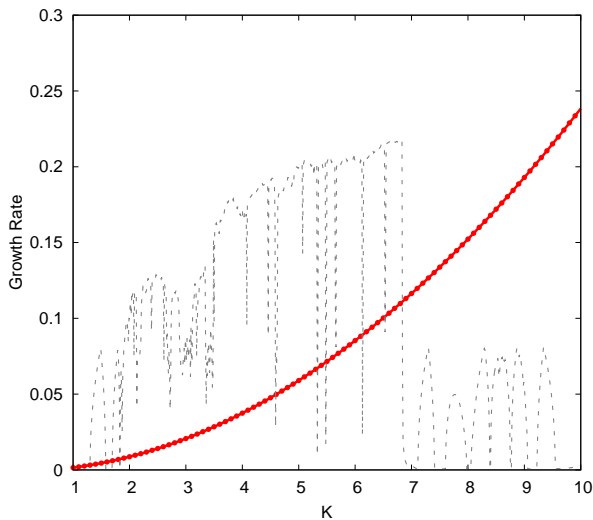}
    \caption{(Color online) The (red) dark gray solid line with circles shows the growth rate of the OTOC of Fig. \ref{fig:fig1} between $t=0$ and $t=1$, 
    as a function of $K$. The gray dashed line displays the average maximum Lyapunov exponent as a function of the same parameter of the classical system (in all cases $\gamma=0.20$).}
    \label{fig:fig2}
\end{figure}

In Fig. \ref{fig:fig1} we observe the OTOC evolution up to a time $ t=15$ for different values of $K$. In all cases we have taken a value of $ \hbar_{\text{eff} } = 0.031 $ and $ \gamma = 0.20 $. The OTOC initially grows very fast and then decays exponentially (log scale in the inset). In chaotic closed systems the OTOC growth has been found to be 
exponential with a rate given by the Lyapunov exponent of its classical counterpart. Interestingly, in our case we do not find that kind of growth, though given its extremely fast nature we cannot be conclusive about a general behavior. In Fig. \ref{fig:fig2} we compare the maximum Lyapunov exponent $\expval{l_{max} }_{\tiny{M}}$, averaged over $M$ initial conditions, $( q_{ 0 }, p_{ 0 } ) $ distributed with uniform probability in the region $[0, 2\pi] \times [\-\pi , \pi]$.

\begin{figure}
    \centering
    \includegraphics[width=7.0cm, angle=-90]{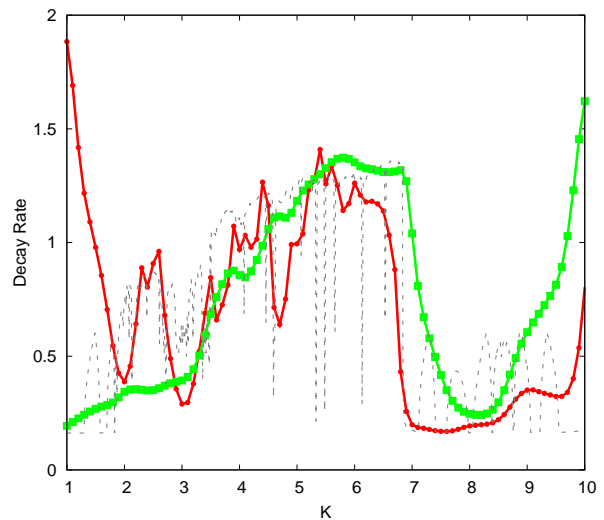}
    \caption{(Color online) The (red) dark gray solid line with circles shows the decay rate of the OTOC of Fig. \ref{fig:fig1} between $t=5$ and $t=100$, as a function of $K$. The (green) light gray solid line with squares corresponds to the quantum IPR values, and the dashed gray line to the average maximum Lyapunov exponent of the classical system (in all cases $\gamma=0.200$).}
    \label{fig:fig3}
\end{figure}

\begin{figure}
    \centering
    \includegraphics[width=7.0cm, angle=-90]{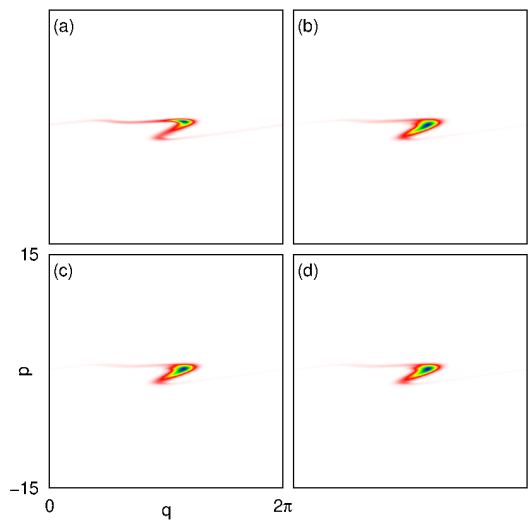}
    \caption{(Color online) Phase space representation of the evolved initial state for times: a) $t=1$, b) $t=3$, c) $t=8$ and d) $t=20$. Lower to higher values of the distributions go from white to (rainbow colors) darker grays. In this case $K=1.10$, $ \hbar_{\text{eff}} = 0.031 $, $ \gamma = 0.200 $ and $ N = 1024 $. }
    \label{fig:fig4}
\end{figure}

\begin{figure}
    \centering
    \includegraphics[width=7.0cm, angle=-90]{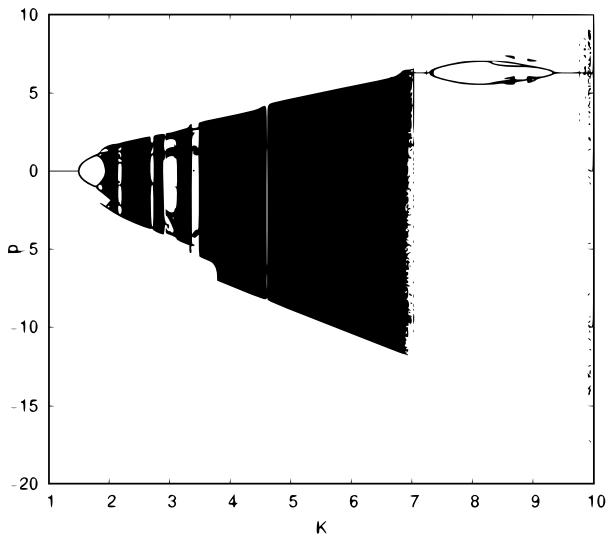}
    \caption{Bifurcation diagram for the classical system given by Eq. \eqref{eq:krmap} for $ \gamma=0.200$.  }
    \label{fig:fig5}
\end{figure}

We have calculated the OTOC decay rate by fitting $\ln({\rm OTOC})$ for values of $K$ from $1$ to $10$ and compared it with the maximum average Lyapunov exponent $l_{max} = \expval{ l^{(i)}_{max} }_{ \tiny{ M } } $ and the IPR defined as
\[ {\rm IPR} = \frac{(\sum{\rho_{ii}^2)^{-1}}}{N}\]
with $ \rho_{ii}$ being the diagonal elements of the density matrix in the momentum basis. The results are shown in Fig. \ref{fig:fig3} where 
the Lyapunov exponent and IPR were rescaled according to $ \Tilde{l}_{ max } = 0.55 \, l_{max} + 0.605$ and $ \Tilde{\rm IPR} = 3.5 \, {\rm IPR}$. The horizontal line at $0.605$ corresponds to zero value for the Lyapunov exponent, i.e. the border between the chaotic and regular regimes. First, we can observe that for $K>2$ the OTOC decays at a lower rate in regular regions, which in turn correspond to a Lyapunov exponent of less than $0.605$. However, for $K<2$ we see that the behavior of the OTOC disagrees with that of the Lyapunov exponent. This can be explained given that the strength of the kick is not enough for the system to explore a significant portion of the phase space and therefore the OTOC cannot account for enough scrambling, decaying before that can happen due to dissipation. This is clearly seen by looking at the evolution of the Husimi functions for $K=1. 10$ displayed in Fig. \ref{fig:fig4}. At larger values of $K$, a much greater region of phase space is explored, and the OTOC is able to capture the interplay between scrambling and decoherence. 
In the region $2<K<7$ the system is mainly chaotic, with specific $K$ values for which regularity is recovered. This can be learned from the Lyapunov exponent and bifurcation diagram of  Fig. \ref{fig:fig5}. The larger regular regions in this portion of the parameter space are detected by the OTOC as can be seen by the decay rate abrupt changes.

On the other hand, for $ K>7 $, the OTOC detects very well the beginning and ending of the largest regular region, in agreement with what it is shown by the Lyapunov exponent and the classical bifurcation diagram. In contrast, the IPR fails to sharply account for both limits, reflecting an effectively smaller regular window. This is because the IPR relies only on the equilibrium state distribution, which suffers from {\em parametric tunneling} \cite{parametric}. The consequences for our understanding of this phenomenon arising from this discrepancy will be explored in subsequent work. We underline here that the OTOC behavior remains essentially unchanged when considering a thermal instead of a coherent state for the calculations.

\begin{figure}
    \hspace{4cm}
    \includegraphics[width=7.0cm, angle=-90]{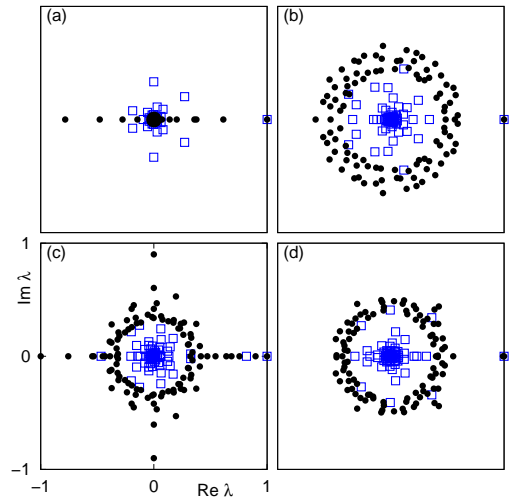}
    \caption{(Color online) 100 largest eigenvalues of the quantum superoperator $e^{\Lambda}$ and of the Perron-Frobenius operator without Gaussian noise, for (a) $K=1.10$, (b) $K=5.40$, (c) $K=8.20$ and (d) $K=10.00$. (Blue) gray squares correspond to the quantum model, while black dots to the classical one. In this case, $\hbar^{\text{(PF)}}_{\text{eff}} = \hbar_{\text{eff}}=0.150$, $\gamma=0.200$ and $N=512$. }
    \label{fig:fig6}
\end{figure}

\begin{figure}
    \centering
    \includegraphics[width=7.0cm, angle=-90]{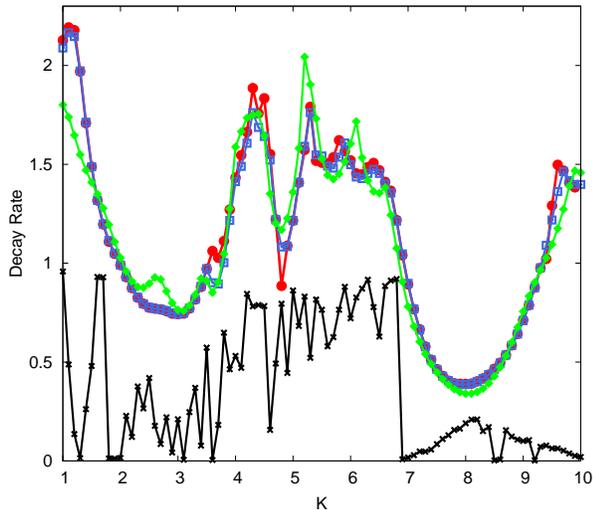}
    \caption{(Color online) The (red) dark gray solid line with circles shows the decay rate of the OTOC between $t=5$ and $t=100$, as a function of $K$. The (blue) gray solid line with empty squares represent $-2\ln{(|\lambda_{1}|)}$ for the eigenvalue $\lambda_{1}$ of the quantum evolution operator, and the (green) light gray solid line with diamonds and black solid line with crosses correspond to the same quantity but for the classical system with and without Gaussian noise, respectively. In all cases $\hbar^{\text{(PF)}}_{\text{eff}}=\hbar_{\text{eff}} = 0.150 $, $ \gamma = 0.200 $ and $ N = 512 $. }
    \label{fig:fig7}
\end{figure}

In the classical system, the Perron-Frobenius operator determines the evolution of Liouville distributions, while in its quantum counterpart this is accomplished by the Lindblad operator $\mathcal{L}$ acting on the density matrix as $\rho_{t+1} = \exp(\mathcal{L}) \rho_{t} = \Lambda \rho_{t}$. Both have an eigenvalue $\lambda_{0} = 1$, whose eigenvector is the attractor to which the initial conditions decay when $t\rightarrow \infty$. The other leading eigenvalues of modulus less than one prescribe the decays. At the classical level, we use Ulam's method to obtain an approximation of the Perron-Frobenius operator from the discretization of the system into cells of size $\hbar^{\text{(PF)}}_{\text{eff}}$. Fig. \ref{fig:fig6} shows the $100$ largest quantum and classical eigenvalues for different values of $K$, clear differences between the spectra can be seen (we have used $ \hbar^{\text{(PF)}}_{\text{eff}}=\hbar_{\text{eff}}=0.150$ to reduce the basis size in which the Perron-Frobenius operator is discretized since this is a computationally demanding task). The spectral gap is given by $ \lambda_{1}$ the decaying eigenvalue with the largest modulus. Fig. \ref{fig:fig7} shows the OTOC decay rate together with $-2\ln(| \lambda_{1}|)$ from both the quantum and classical spectra as a function of $K$. The agreement between the OTOC and the quantum spectral decay is generally very good, even for low $K$. There are values at which the differences are noticeable though, for example at $K \approx 3.5$. This is due to the fact that in this parameter region chaotic and regular behavior coexist as observed in the bifurcation diagram of Fig. \ref{fig:fig5}. On the other hand, the decay rate associated to the classical spectrum behaves very differently, implying that the OTOC behavior is not directly predicted by the Ruelle-Pollicott resonances in a simple way. In \cite{carlo2019} these discrepancies proved to be drastically reduced by adding Gaussian noise $\xi$ (with $\expval{\xi}=0$ and $\expval{\xi}^{2} = \hbar_{\text{eff}}$) only to the classical system. If we do this, the disagreement between the classical and the OTOC decay rates disappear and the quantum to classical correspondence is recovered.

Summarizing our results, we have found that in the quantum DMKRM, a paradigmatic dissipative system, the OTOC grows very fast at short times and then decays in an exponential way at a rate essentially given by the (rescaled) Lyapunov exponent of the classical counterpart. This decay depends on the dissipation and forcing strengths $\gamma$ and $K$, which determine the chaotic or regular nature of the dynamics. Moreover, the OTOC revealed itself as much more sensitive to the dynamical regime type when compared with other complexity measure based on the eigenstates distributions, like the very frequently used IPR. However, when the distributions are not able to explore a meaningful portion of phase space the scrambling power of the system is not sufficiently captured by the OTOC, and consequently it becomes a poor complexity detector as happens for $K<2$ in our model. All this points towards the fact that capturing the interplay between scrambling and dissipation is of the essence at the time to characterize the complexity of quantum dissipative systems. In the vast majority of the dynamical scenarios we have looked into, studying the OTOC showed as a very suitable way to do it.

On the other hand, the largest decaying eigenvalue $\lambda_{1}$ determines in almost the entire range of $K$ the decay rate of the OTOC, except for a few small regions to be analyzed in more detail in future work. We recall that changes in the classical eigenvalue spectrum indicate the loss or gain of stability of the attractor \cite{tantet2018resonances}, therefore the OTOC could be a good way to extend this result to the quantum realm. Finally, by adding $\hbar_{\text{eff}}$-sized Gaussian noise only to the classical system, we recover the OTOC decay rate being this a nice example of the quantum to classical correspondence principle of quantum dissipative systems \cite{carlo2012quantum, carlo2019}

Financial support from CONICET is gratefully acknowledged.



\end{document}